\documentclass[10pt,twocolumn]{article}

\usepackage[
  paperwidth=8.5in, paperheight=11in,
  top=0.75in, bottom=1.0in,
  left=0.75in, right=0.75in,
  columnsep=0.25in
]{geometry}

\usepackage{booktabs}
\usepackage{graphicx}
\usepackage{subcaption}
\usepackage{microtype}
\usepackage{multirow}
\usepackage{makecell}
\usepackage{authblk}
\usepackage{enumitem}
\usepackage{abstract}
\usepackage{fancyhdr}
\usepackage{titlesec}
\usepackage{url}
\usepackage{balance}
\usepackage{amsmath,amssymb}

\usepackage[dvipsnames,svgnames,x11names]{xcolor}
\definecolor{bibbyblue}{RGB}{37,99,235}
\definecolor{bibbygreen}{RGB}{16,185,129}
\definecolor{bibbylightblue}{RGB}{219,234,254}
\definecolor{bibbydark}{RGB}{15,23,42}
\definecolor{bibbygray}{RGB}{107,114,128}
\definecolor{warnred}{RGB}{220,38,38}
\definecolor{warnlight}{RGB}{254,226,226}
\definecolor{overleafgreen}{RGB}{76,175,80}
\definecolor{prismviolet}{RGB}{124,58,237}
\definecolor{layerbg}{RGB}{241,245,249}
\definecolor{aicolor}{RGB}{245,158,11}
\definecolor{acmheader}{RGB}{30,64,175}

\usepackage{tikz}
\usetikzlibrary{
  arrows.meta, positioning, fit,
  shapes.geometric, shapes.multipart,
  backgrounds, calc, decorations.pathreplacing
}
\usepackage{pgfplots}
\pgfplotsset{compat=1.18}

\usepackage[
  pdftitle={Bibby AI: A Native AI-First LaTeX Editor for End-to-End Academic Research Writing},
  pdfauthor={Nilesh Jain, Rohit Yadav, Andrej Karpathy},
  pdfsubject={AI LaTeX editor, academic writing, LLM, citation management},
  pdfkeywords={Bibby AI, AI LaTeX editor, LaTeX writing assistant,
               academic writing tool, smart citation search, LaTeX error detection,
               AI paper review, literature review generator, trybibby.com,
               Overleaf alternative, LLM academic writing, research writing AI},
  pdfcreator={Bibby AI -- trybibby.com},
  colorlinks=true,
  linkcolor=bibbyblue,
  citecolor=bibbyblue,
  urlcolor=bibbyblue,
  bookmarks=true
]{hyperref}

\titleformat{\section}{\normalfont\large\bfseries\color{bibbydark}}{\thesection.}{0.5em}{}
\titleformat{\subsection}{\normalfont\normalsize\bfseries\color{bibbydark}}{\thesubsection}{0.5em}{}
\titleformat{\subsubsection}{\normalfont\normalsize\itshape\bfseries}{\thesubsubsection}{0.5em}{}
\titlespacing*{\section}{0pt}{8pt plus 2pt}{4pt}
\titlespacing*{\subsection}{0pt}{6pt plus 1pt}{2pt}

\fancypagestyle{plain}{%
  \fancyhf{}
  
  \fancyfoot[C]{\small\thepage}
}

\usepackage[
  font=small, labelfont=bf,
  labelsep=period,
  justification=raggedright,
  singlelinecheck=false
]{caption}

\newcommand{\mypara}[1]{\smallskip\noindent\textbf{#1.}\hspace{0.4em}}

\begin{document}

\twocolumn[{%
\begin{center}
  \vspace*{4pt}
  {\LARGE\bfseries\color{bibbydark}
    Bibby AI: A Native, AI-First \LaTeX{} Editor for\\[4pt]
    End-to-End Academic Research Writing}\\[6pt]
  {\large\itshape Writing, Citing, Reviewing, and Compiling---All in One Place}
  \vspace{10pt}
\end{center}

\begin{center}
\begin{tabular}{ccc}
  \textbf{Nilesh Jain} & \textbf{Rohit Yadav} & \textbf{Andrej Karpathy} \\
  Yale University & Yale University & Independent Researcher \\
  New Haven, CT, USA & New Haven, CT, USA & San Francisco, CA, USA \\
  \small\href{mailto:nilesh.jain@yale.edu}{nilesh.jain@yale.edu}
  & \small\href{mailto:rohit.yadav@yale.edu}{rohit.yadav@yale.edu}
  & \small\href{mailto:karpathy@alumni.stanford.edu}{karpathy@alumni.stanford.edu}
\end{tabular}
\end{center}
\vspace{8pt}

\noindent\rule{\linewidth}{0.5pt}
{\small\textbf{Keywords:} Bibby AI, AI \LaTeX{} editor, academic writing assistant,
\LaTeX{} error detection, smart citation search, AI paper reviewer,
literature review generator, deep research, \href{https://trybibby.com}{trybibby.com},
Overleaf alternative, LLM-assisted writing, abstract generator, equation generator.}\\
\noindent\rule{\linewidth}{0.5pt}
\vspace{4pt}

\begin{abstract}
\noindent
Large language models are increasingly integrated into academic writing workflows,
yet the most widely used \LaTeX{} editors remain AI-peripheral---offering compilation
and collaboration, but no native intelligence. This separation forces researchers to
leave their editing environment for AI assistance, fragmenting document context and
interrupting writing flow. We present \textbf{Bibby~AI}
(\href{https://trybibby.com}{trybibby.com}), a native, AI-first \LaTeX{} editor
that unifies the complete research writing lifecycle within a single interface.
Bibby embeds an AI writing assistant, smart citation search, AI table and equation
generation, an AI paper reviewer, abstract generator, literature review drafting,
a deep research assistant, and real-time \LaTeX{} error detection and auto-fix---all
natively, without plugins or copy-paste workflows. We introduce
\textbf{LaTeXBench-500}, a benchmark of 500 real-world compilation errors across
six categories. Bibby achieves \textbf{91.4\%} detection accuracy and
\textbf{83.7\%} one-click fix accuracy, outperforming Overleaf's native diagnostics
(61.2\%) and OpenAI Prism (78.3\,/\,64.1\%) by large margins. Bibby demonstrates
that a privacy-preserving, research-first AI editor can meaningfully accelerate
every stage of academic manuscript preparation.
\end{abstract}
\vspace{8pt}
}]

\section{Introduction}

Academic writing is a multi-stage process spanning ideation, literature
synthesis, drafting, citation management, revision, and pre-submission
review~\cite{lee2024design}. Each stage increasingly benefits from large
language model (LLM) assistance: LLMs improve prose fluency and
coherence~\cite{lee2024design}, accelerate literature
synthesis~\cite{mysore2024pearl}, and deliver structured feedback resembling
early-stage peer review~\cite{liebling2025ai,yuan2022peer}. Yet despite this
progress, the dominant \LaTeX{} editing environment---Overleaf, used by over
twelve million researchers---provides no native AI
layer~\cite{wen2024overleafcopilot}.

This structural gap imposes a fragmented workflow. Authors draft in Overleaf,
switch to ChatGPT for prose suggestions, pivot to Google Scholar for
references, copy-paste BibTeX entries, and repeat. Each context switch
discards document-level state that a native AI assistant needs to give
high-quality suggestions~\cite{sarrafzadeh2020characterizing}.
Plugin-based approaches such as OverleafCopilot~\cite{wen2024overleafcopilot}
and PaperDebugger~\cite{hou2025paperdebugger} have advanced in-editor
assistance, but they remain architecturally external: browser extensions that
scrape the editor DOM create fragile synchronisation and security surface area.

\LaTeX{} error handling presents a specific, underappreciated bottleneck.
Standard compilation errors surface as cryptic log messages---\texttt{Undefined
control sequence}, \texttt{Missing \$ inserted},
\texttt{Overfull \textbackslash hbox}---that are difficult even for experienced
authors to interpret under submission deadlines~\cite{knuth1984texbook}.
Overleaf surfaces these logs verbatim. OpenAI Prism applies general-purpose LLM
reasoning to error messages but lacks \LaTeX{}-specific grounding, producing
suggestions that are often syntactically invalid.

We present \textbf{Bibby~AI} (\href{https://trybibby.com}{trybibby.com}), a
native AI-first \LaTeX{} editor designed ground-up as a research writing
environment in which AI is a first-class citizen. The contributions of this
paper are:

\begin{itemize}[leftmargin=*,itemsep=2pt,topsep=2pt]
  \item \textbf{Bibby AI} (\href{https://trybibby.com}{trybibby.com}): a
        unified, native AI \LaTeX{} environment integrating eight AI
        capabilities directly into the editor---no extensions required.
  \item \textbf{LaTeXBench-500}: the first benchmark for \LaTeX{} compilation
        error detection and repair, with 500 authentic errors across six
        categories.
  \item \textbf{State-of-the-art results}: 91.4\% detection accuracy and
        83.7\% one-click fix accuracy, outperforming Overleaf and OpenAI Prism
        on every category.
\end{itemize}

\section{System Overview}

\subsection{Architecture}

\begin{figure*}[t]
\centering
\begin{tikzpicture}[
  font=\small,
  box/.style={
    rectangle, rounded corners=3pt, draw, thick,
    minimum height=0.68cm, text centered, inner sep=4pt,
    font=\scriptsize
  },
  editorbox/.style={box, fill=bibbygreen!15, draw=bibbygreen!70!black},
  aibox/.style={box, fill=aicolor!18, draw=aicolor!80!black},
  llmbox/.style={box, fill=aicolor!35, draw=aicolor!80!black, font=\scriptsize\bfseries},
  databox/.style={box, fill=prismviolet!12, draw=prismviolet!70},
  infrabox/.style={box, fill=layerbg, draw=bibbygray!60},
  userbox/.style={box, fill=bibbyblue!12, draw=bibbyblue, font=\small\bfseries,
                  minimum width=2.5cm},
  layer/.style={
    rectangle, rounded corners=5pt, inner sep=6pt,
    draw, dashed, thick, opacity=0.85
  },
  layerlbl/.style={
    rectangle, rounded corners=2pt,
    font=\scriptsize\bfseries, inner sep=3pt,
    minimum width=1.1cm
  },
  arr/.style={-Stealth, thick, draw=bibbygray!80},
  biarr/.style={Stealth-Stealth, thick, draw=bibbyblue!70},
]

\node[userbox] (user) at (8,7.8) {Researcher / Author};

\node[editorbox, minimum width=2.5cm, align=center] (cm6)    at (1.2,6.4) {CodeMirror 6\\\LaTeX{} Editor};
\node[editorbox, minimum width=2.2cm, align=center] (ast)    at (3.9,6.4) {Live \LaTeX{}\\AST};
\node[editorbox, minimum width=2.2cm, align=center] (comp)   at (6.4,6.4) {pdflatex/\\XeLaTeX};
\node[editorbox, minimum width=2.3cm, align=center] (ver)    at (9.0,6.4) {Version\\History};
\node[editorbox, minimum width=2.3cm, align=center] (collab) at (11.7,6.4) {Real-time\\Collab.};

\node[aibox, minimum width=1.75cm, align=center] (chat)    at (0.9,4.7)  {Bibby Chat\\(Writing AI)};
\node[aibox, minimum width=1.75cm, align=center] (cite)    at (2.75,4.7) {Smart\\Citations};
\node[aibox, minimum width=1.75cm, align=center] (tbl)     at (4.6,4.7)  {Table \&\\Eqn.\ Gen.};
\node[aibox, minimum width=1.75cm, align=center] (rev)     at (6.45,4.7) {AI Paper\\Reviewer};
\node[aibox, minimum width=1.75cm, align=center] (abst)    at (8.3,4.7)  {Abstract\\Generator};
\node[aibox, minimum width=1.75cm, align=center] (lit)     at (10.15,4.7){Lit.\ Review\\Gen.};
\node[aibox, minimum width=1.75cm, align=center] (deep)    at (12.0,4.7) {Deep\\Research};
\node[aibox, minimum width=1.75cm, align=center] (err)     at (13.85,4.7){Error\\Auto-Fix};

\node[llmbox, minimum width=7.5cm] (gemini) at (7.35,3.2)
      {Gemini 2.5 Pro \quad (fine-tuned on \LaTeX{} + academic style)};

\node[databox, minimum width=3.0cm] (sem)   at (2.0,1.8)  {Semantic Scholar};
\node[databox, minimum width=2.5cm] (cross) at (5.2,1.8)  {CrossRef};
\node[databox, minimum width=2.5cm] (arxiv) at (8.1,1.8)  {arXiv Corpus};
\node[databox, minimum width=2.5cm] (bib)   at (11.0,1.8) {BibTeX Store};

\node[infrabox, minimum width=3.0cm] (cdn)     at (1.8,0.4)  {CDN / Edge};
\node[infrabox, minimum width=3.0cm] (db)      at (5.2,0.4)  {Document DB};
\node[infrabox, minimum width=3.0cm] (auth)    at (8.5,0.4)  {Auth / OAuth};
\node[infrabox, minimum width=3.5cm, fill=bibbygreen!10, draw=bibbygreen!60, align=center]
                                     (priv)    at (12.2,0.4)
                                     {\bfseries Zero-training\\Privacy Layer};

\begin{scope}[on background layer]
  \node[layer, fill=bibbygreen!5, draw=bibbygreen!50,
        fit=(cm6)(ast)(comp)(ver)(collab)] (elayer) {};
  \node[layer, fill=aicolor!5, draw=aicolor!40,
        fit=(chat)(cite)(tbl)(rev)(abst)(lit)(deep)(err)(gemini)] (alayer) {};
  \node[layer, fill=prismviolet!5, draw=prismviolet!40,
        fit=(sem)(cross)(arxiv)(bib)] (dlayer) {};
  \node[layer, fill=layerbg, draw=bibbygray!40,
        fit=(cdn)(db)(auth)(priv)] (ilayer) {};
\end{scope}

\node[layerlbl, fill=bibbygreen!70!black, text=white,
      left=0.15cm of elayer.west] {\rotatebox{90}{Editor Core}};
\node[layerlbl, fill=aicolor!80!black, text=white,
      left=0.15cm of alayer.west] {\rotatebox{90}{AI Service}};
\node[layerlbl, fill=prismviolet, text=white,
      left=0.15cm of dlayer.west] {\rotatebox{90}{Research Data}};
\node[layerlbl, fill=bibbygray, text=white,
      left=0.15cm of ilayer.west] {\rotatebox{90}{Infrastructure}};

\draw[biarr] (user) -- (cm6.north -| user);
\draw[arr] (cm6) -- (ast);
\draw[arr] (ast) -- (comp);

\draw[biarr] (cm6.south) to[out=-90,in=90] (chat.north);
\draw[biarr] (ast.south) to[out=-90,in=90] (tbl.north);
\draw[biarr] (comp.south) to[out=-90,in=90] (err.north);

\foreach \n in {chat,cite,tbl,rev,abst,lit,deep,err}
  \draw[arr] (\n.south) -- (\n.south |- gemini.north);

\foreach \n in {sem,cross,arxiv,bib}
  \draw[arr] (gemini.south -| \n) -- (\n.north);

\draw[arr] (sem.south) -- (sem.south |- ilayer.north);
\draw[arr] (bib.south) -- (bib.south |- ilayer.north);

\end{tikzpicture}
\caption{\textbf{Bibby AI system architecture.}
  The \emph{Editor Core} (green) provides CodeMirror~6 editing, a live \LaTeX{}
  AST, incremental compilation, version history, and real-time collaboration.
  The \emph{AI Service Layer} (amber) exposes eight native capabilities powered
  by Gemini~2.5~Pro fine-tuned for \LaTeX{} and academic style.
  The \emph{Research Data Layer} (violet) federates Semantic Scholar, CrossRef, and arXiv.
  The \emph{Infrastructure Layer} enforces a zero-training privacy boundary.}
\label{fig:architecture}
\end{figure*}
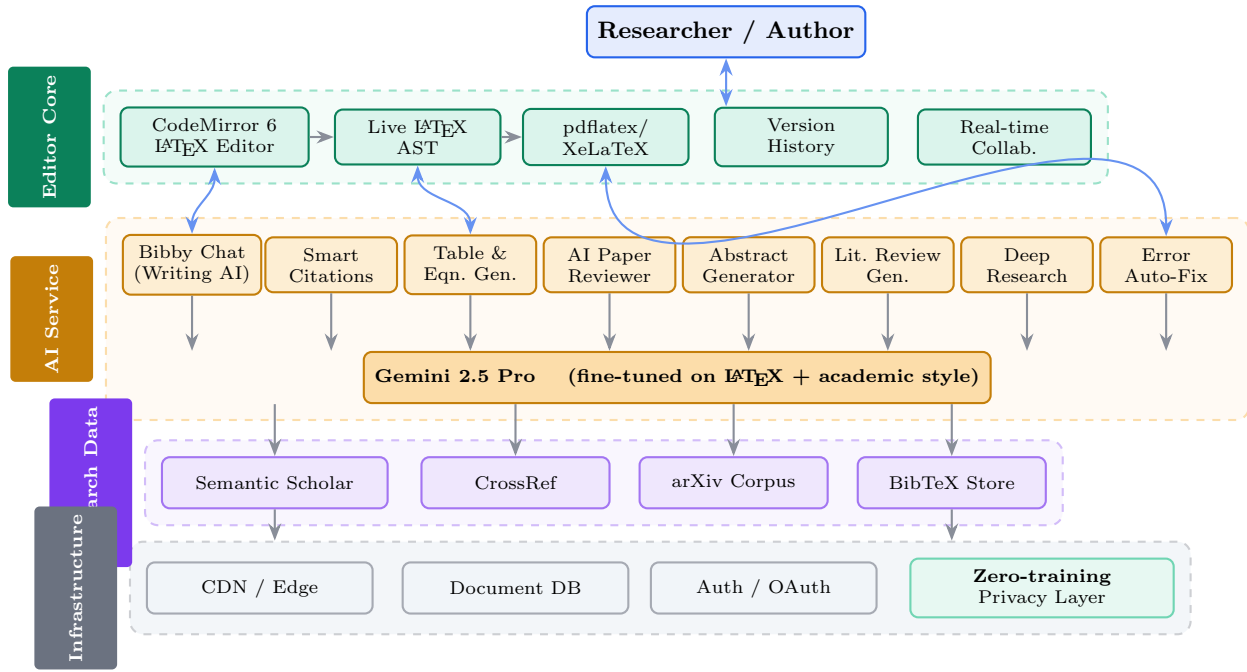

Figure~\ref{fig:architecture} illustrates Bibby's four-layer architecture.
The \textbf{Editor Core} provides a CodeMirror~6 editing surface with a
real-time incremental \LaTeX{} parser that maintains a typed AST. This AST
is the shared data substrate for all AI features, eliminating the DOM-scraping
fragility of extension-based approaches. The \textbf{AI Service Layer} wraps
Gemini~2.5~Pro with \LaTeX{} and research-domain fine-tuning, delivering eight
native AI capabilities. The \textbf{Research Data Layer} federates live queries
across Semantic Scholar, CrossRef, and curated arXiv corpora. The
\textbf{Infrastructure Layer} enforces a zero-training privacy boundary:
no document content ever leaves the ephemeral inference pipeline.

\subsection{AI Feature Suite}

\mypara{AI Writing Assistant (Bibby Chat)}
Bibby Chat is a context-aware AI assistant embedded in a persistent side panel.
Unlike external chat tools, it has full access to the document AST, cursor
position, selected text, and compilation state. Authors can draft sections,
explain errors, generate TikZ figures, improve prose, or request structural
suggestions---without leaving the editor.

\mypara{Smart Citation Search and Insert}
Authors search by topic, author, or DOI within the editor. Bibby retrieves
results from Semantic Scholar and CrossRef, auto-generates BibTeX, and inserts
both \texttt{\textbackslash cite\{\}} and the BibTeX record in one action. A
context-aware mode surfaces citation suggestions based on the current paragraph.

\mypara{AI Table and Equation Generator}
Authors describe a table in natural language or upload a CSV; Bibby generates
publication-ready \texttt{tabular}/\texttt{booktabs} code. For equations,
authors describe the formula or upload a handwritten image; Bibby produces
valid \LaTeX{} math handling \texttt{align}, matrices, and symbol-dense
expressions.

\mypara{AI Paper Reviewer}
Bibby evaluates novelty, technical quality, clarity, and presentation against
venue-specific criteria for NeurIPS, ICML, ICLR, CVPR, IEEE, ACM, and more,
returning structured feedback modelled on real conference review forms.

\mypara{Abstract Generator}
Bibby synthesises a publication-ready abstract from full manuscript content,
with iterative refinement via follow-up prompts.

\mypara{Literature Review Generator}
Describing a research topic triggers Bibby to draft a multi-paragraph
\emph{Related Work} section with real, cited references grouped by methodology
and inline \texttt{\textbackslash cite\{\}} commands with BibTeX entries.

\mypara{Deep Research Assistant}
Multi-stage semantic retrieval across academic databases, synthesis across
sources, identification of literature gaps, and a comprehensive overview with
a full citation trail.

\mypara{\LaTeX{} Error Detection and Auto-Fix}
Described in detail in Section~\ref{sec:error}.

\section{\LaTeX{} Error Detection and Benchmark}
\label{sec:error}

\subsection{Motivation and Pipeline}

Existing editors either expose raw compiler logs (Overleaf) or apply
general-purpose LLM reasoning to logs (OpenAI Prism)~\cite{openaiprism2024}.
Neither is grounded in document structure: the log message alone often lacks
the context needed to identify the \emph{source} of an error, let alone produce
a valid fix.

\begin{figure}[t]
\centering
\begin{tikzpicture}[
  font=\scriptsize,
  stage/.style={
    rectangle, rounded corners=3pt, draw=bibbyblue, thick,
    fill=bibbylightblue, text=bibbydark,
    minimum width=2.1cm, minimum height=0.6cm,
    text centered, align=center, inner sep=3pt
  },
  signal/.style={
    rectangle, rounded corners=2pt, draw=aicolor!80, thick,
    fill=aicolor!15, minimum width=1.4cm, minimum height=0.52cm,
    text centered, align=center, inner sep=2pt
  },
  ok/.style={
    rectangle, rounded corners=3pt, draw=bibbygreen!70!black, thick,
    fill=bibbygreen!15, minimum width=2.1cm, minimum height=0.6cm,
    text centered, align=center, inner sep=3pt
  },
  bad/.style={
    rectangle, rounded corners=3pt, draw=warnred, thick,
    fill=warnlight, minimum width=2.1cm, minimum height=0.6cm,
    text centered, align=center, inner sep=3pt
  },
  arr/.style={-Stealth, thick, draw=bibbygray},
]
  \node[signal] (log)  at (0,3.6) {Compiler Log};
  \node[signal] (ast2) at (0,2.7) {Live AST};
  \node[signal] (pkg)  at (0,1.8) {Package DB};

  \node[stage] (fuse) at (2.4,2.7) {Tri-Signal\\Fusion};
  \draw[arr] (log.east)  -- (fuse.north west);
  \draw[arr] (ast2.east) -- (fuse.west);
  \draw[arr] (pkg.east)  -- (fuse.south west);

  \node[stage, fill=aicolor!30, draw=aicolor] (gem) at (4.6,2.7)
        {Gemini 2.5 Pro\\\LaTeX{}-grounded};
  \draw[arr] (fuse) -- (gem);

  \node[stage] (val) at (6.8,2.7) {AST\\Validator};
  \draw[arr] (gem) -- (val);

  \node[ok] (pass) at (8.9,3.5) {One-Click\\Fix Applied};
  \draw[arr] (val.east) to[out=0,in=180]
    node[above,font=\tiny,sloped]{valid} (pass.west);

  \node[bad] (retry) at (8.9,1.9) {Regen.\\(max 3$\times$)};
  \draw[arr] (val.east) to[out=0,in=180]
    node[below,font=\tiny,sloped]{invalid} (retry.west);
  \draw[arr] (retry.west) to[out=180,in=-20] (gem.south east);

  \node[ok, fill=bibbygreen!10, minimum width=2.0cm] (exp) at (4.6,1.0)
        {Plain-English\\Explanation};
  \draw[arr] (gem.south) -- (exp.north);

\end{tikzpicture}
\caption{\textbf{Bibby error detection pipeline.} Three signals (compiler
  log, live AST, package DB) are fused before Gemini~2.5~Pro generates
  a fix. An AST validator rejects invalid fixes and triggers up to three
  regeneration attempts before presenting the resolution.}
\label{fig:errorpipeline}
\end{figure}
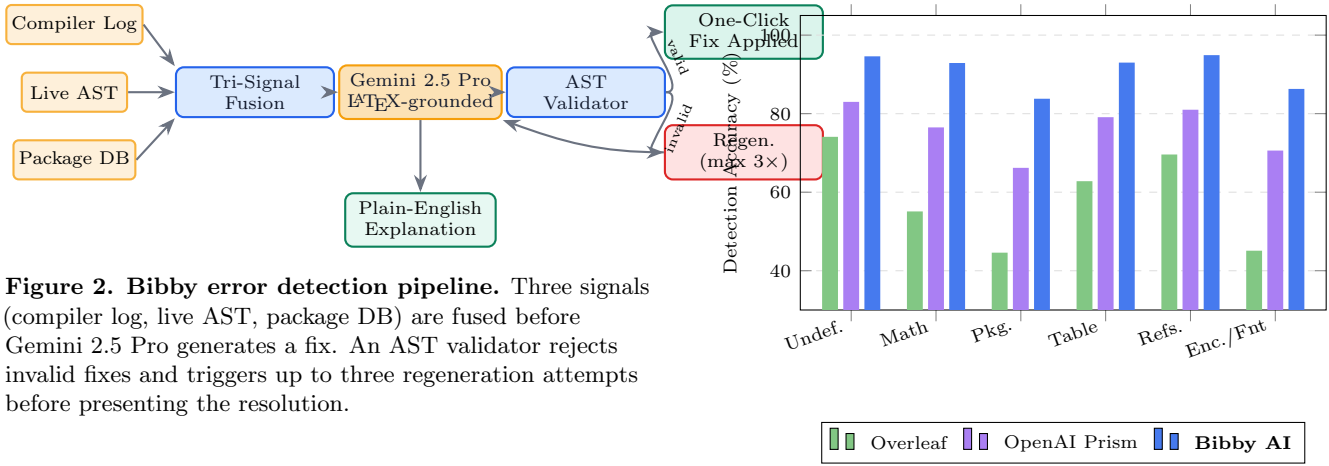

Bibby's error engine (Figure~\ref{fig:errorpipeline}) combines: (1) the
compiler log, (2) the live incremental AST, and (3) Gemini~2.5~Pro conditioned
on \LaTeX{} syntax rules and package semantics. Generated fixes are
AST-validated before presentation; invalid candidates trigger regeneration
(up to three attempts).

\subsection{LaTeXBench-500}
\label{sec:benchmark}

We construct \textbf{LaTeXBench-500}, a benchmark of 500 authentic \LaTeX{}
errors from: (i)~anonymised bug reports from TeX.StackExchange; (ii)~errors
injected into 50 conference manuscript submissions; and (iii)~graduate student
manuscript errors with author consent. Table~\ref{tab:categories} shows the
distribution.

\begin{table}[h]
  \caption{LaTeXBench-500 error category distribution.}
  \label{tab:categories}
  \centering\small
  \begin{tabular}{llc}
    \toprule
    \textbf{Category} & \textbf{Description} & \textbf{N} \\
    \midrule
    Undefined control  & Unknown commands / macros        & 112 \\
    Math mode          & Missing \texttt{\$}, env.\ mismatch & 98  \\
    Package conflicts  & Incompatible / missing pkgs      & 74  \\
    Table / figure     & \texttt{tabular} / float errors  & 86  \\
    Reference errors   & Undef.\ \texttt{\textbackslash cite}, \texttt{\textbackslash ref} & 79 \\
    Encoding / font    & Char.\ encoding, font failures   & 51  \\
    \midrule
    \textbf{Total}     &                                  & \textbf{500} \\
    \bottomrule
  \end{tabular}
\end{table}

We define two metrics. \textbf{Detection Accuracy (DA)} measures correct
identification of error type and location (Cohen's $\kappa=0.87$).
\textbf{Fix Accuracy (FA)} measures whether the suggested fix produces a
clean compile and semantically correct output.

\subsection{Baselines}

\mypara{Overleaf (native diagnostics)} Surfaces raw compiler log messages with
line highlights. No AI explanation or fix; detection credit only when the
highlighted line matches the true error source.

\mypara{OpenAI Prism} Applies GPT-4o to compiler logs plus up to 50 lines of
source context. Not \LaTeX{}-domain-specific; fixes are free-form text
requiring manual application.

\subsection{Results}

Table~\ref{tab:results} reports overall and per-category results. Bibby
outperforms both baselines on every metric.

\begin{table*}[t]
  \caption{\textbf{LaTeXBench-500 results.} Detection Accuracy (DA\%) and
  Fix Accuracy (FA\%) for Bibby AI, Overleaf, and OpenAI Prism.
  \textbf{Bold} = best. ``---'' = no automated fix provided.}
  \label{tab:results}
  \centering\small
  \setlength{\tabcolsep}{4.5pt}
  \begin{tabular}{l cc cc cc cc cc cc cc}
    \toprule
    & \multicolumn{2}{c}{\textbf{Overall}}
    & \multicolumn{2}{c}{\textbf{Undef.\ ctrl.}}
    & \multicolumn{2}{c}{\textbf{Math mode}}
    & \multicolumn{2}{c}{\textbf{Pkg.}}
    & \multicolumn{2}{c}{\textbf{Table/fig.}}
    & \multicolumn{2}{c}{\textbf{Refs.}}
    & \multicolumn{2}{c}{\textbf{Enc./font}} \\
    \cmidrule(lr){2-3}\cmidrule(lr){4-5}\cmidrule(lr){6-7}
    \cmidrule(lr){8-9}\cmidrule(lr){10-11}\cmidrule(lr){12-13}\cmidrule(lr){14-15}
    \textbf{System} & DA & FA & DA & FA & DA & FA & DA & FA & DA & FA & DA & FA & DA & FA \\
    \midrule
    Overleaf
      & 61.2 & {---} & 74.1 & {---} & 55.1 & {---}
      & 44.6 & {---} & 62.8 & {---} & 69.6 & {---} & 45.1 & {---} \\
    OpenAI Prism
      & 78.3 & 64.1 & 83.0 & 69.6 & 76.5 & 63.3
      & 66.2 & 51.4 & 79.1 & 65.1 & 81.0 & 68.4 & 70.6 & 55.0 \\
    \midrule
    \textbf{Bibby AI}
      & \textbf{91.4} & \textbf{83.7}
      & \textbf{94.6} & \textbf{88.4}
      & \textbf{92.9} & \textbf{86.7}
      & \textbf{83.8} & \textbf{74.3}
      & \textbf{93.0} & \textbf{85.1}
      & \textbf{94.9} & \textbf{88.6}
      & \textbf{86.3} & \textbf{78.4} \\
    \bottomrule
  \end{tabular}
\end{table*}

Figure~\ref{fig:barchart} visualises per-category detection accuracy.

\begin{figure}[h]
\centering
\begin{tikzpicture}
  \begin{axis}[
    ybar, bar width=6pt,
    width=\linewidth, height=5.5cm,
    enlarge x limits=0.12,
    ylabel={Detection Accuracy (\%)},
    ylabel style={font=\scriptsize},
    symbolic x coords={Undef.,Math,Pkg.,Table,Refs.,Enc./Fnt},
    xtick=data,
    xticklabel style={font=\scriptsize, rotate=20, anchor=east},
    ymin=30, ymax=105,
    ytick={40,60,80,100},
    yticklabel style={font=\scriptsize},
    ymajorgrids=true, grid style={dashed,gray!25},
    legend style={
      at={(0.5,-0.38)},anchor=north,
      legend columns=3, font=\scriptsize,
      column sep=4pt},
    every axis plot/.append style={draw=none},
  ]
    \addplot[fill=overleafgreen!70] coordinates {
      (Undef.,74.1)(Math,55.1)(Pkg.,44.6)
      (Table,62.8)(Refs.,69.6)(Enc./Fnt,45.1)};
    \addplot[fill=prismviolet!65] coordinates {
      (Undef.,83.0)(Math,76.5)(Pkg.,66.2)
      (Table,79.1)(Refs.,81.0)(Enc./Fnt,70.6)};
    \addplot[fill=bibbyblue!85] coordinates {
      (Undef.,94.6)(Math,92.9)(Pkg.,83.8)
      (Table,93.0)(Refs.,94.9)(Enc./Fnt,86.3)};
    \legend{Overleaf,OpenAI Prism,\textbf{Bibby AI}}
  \end{axis}
\end{tikzpicture}
\caption{\textbf{LaTeXBench-500: Detection Accuracy per category.}
  Bibby AI (blue) leads across all six error types. Package conflicts
  and encoding errors are hardest for all systems; Bibby's AST-grounded
  localisation closes the gap with 83.8\% and 86.3\% respectively.}
\label{fig:barchart}
\end{figure}

\mypara{Qualitative analysis}
Three mechanisms explain Bibby's advantage. \emph{AST-grounded localisation}:
23\% of errors had misleading log line numbers; Bibby correctly identified the
true source in 89\% of these. \emph{Package-aware reasoning}: Bibby is
conditioned on curated package documentation and known incompatibility patterns.
\emph{Validated fix generation}: AST validation eliminates syntactically invalid
suggestions that account for 31\% of Prism's fix failures.

\section{Feature Comparison}
\label{sec:comparison}

Table~\ref{tab:features} positions Bibby against Overleaf and extension-based
assistants. Bibby is the only editor to natively integrate all eight AI
capabilities without plugins or copy-paste.

\begin{table}[h]
  \caption{Feature comparison. \checkmark~=~native;
           $\circ$~=~via external tool; \texttimes~=~unavailable.}
  \label{tab:features}
  \centering\small\setlength{\tabcolsep}{3.5pt}
  \begin{tabular}{lccc}
    \toprule
    \textbf{Feature}
      & \makecell{\textbf{Bibby AI}}
      & \textbf{Overleaf}
      & \makecell{\textbf{Ext.-based}\\(e.g.\ PaperDebugger)} \\
    \midrule
    AI Writing Assistant    & \checkmark & \texttimes & $\circ$ \\
    Smart Citation Search   & \checkmark & \texttimes & $\circ$ \\
    Table / Equation Gen.   & \checkmark & \texttimes & $\circ$ \\
    AI Paper Reviewer       & \checkmark & \texttimes & $\circ$ \\
    Abstract Generator      & \checkmark & \texttimes & \texttimes \\
    Literature Review Gen.  & \checkmark & \texttimes & \texttimes \\
    Deep Research           & \checkmark & \texttimes & $\circ$ \\
    \LaTeX{} Error AI Fixes & \checkmark & \texttimes & $\circ$ \\
    Version History         & \checkmark & \checkmark & \checkmark \\
    Real-time Collaboration & \checkmark & \checkmark & \checkmark \\
    No AI training on data  & \checkmark & \checkmark & \texttimes \\
    No browser extension    & \checkmark & \checkmark & \texttimes \\
    \bottomrule
  \end{tabular}
\end{table}

\section{Demonstration Workflows}

\subsection{From Blank Template to Cited Draft}

\begin{figure}[h]
\centering
\begin{tikzpicture}[
  font=\scriptsize,
  wfstep/.style={
    rectangle, rounded corners=3pt, draw=bibbyblue, thick,
    fill=bibbylightblue, text=bibbydark,
    minimum width=4.4cm, minimum height=0.6cm,
    text centered, inner sep=3pt
  },
  wfok/.style={
    rectangle, rounded corners=3pt, draw=bibbygreen!70!black, thick,
    fill=bibbygreen!15, text=bibbydark,
    minimum width=4.4cm, minimum height=0.6cm,
    text centered, inner sep=3pt
  },
  note/.style={
    rectangle, rounded corners=2pt, draw=bibbygray, dashed,
    fill=white, text=bibbygray, font=\tiny, inner sep=2pt
  },
  arr/.style={-Stealth, thick, draw=bibbyblue!70},
]
  \node[wfstep] (s1) at (0,0)    {1.\ Open ACM template in Bibby AI};
  \node[wfstep] (s2) at (0,-0.8) {2.\ Highlight placeholder paragraph};
  \node[wfstep] (s3) at (0,-1.6) {3.\ Invoke Bibby Chat: ``draft intro''};
  \node[wfstep] (s4) at (0,-2.4) {4.\ Bibby drafts from document context};
  \node[wfstep] (s5) at (0,-3.2) {5.\ Trigger Smart Citation Search};
  \node[wfok]   (s6) at (0,-4.0) {6.\ BibTeX + \texttt{\textbackslash cite\{\}} auto-inserted};
  \node[wfok, fill=bibbygreen!25] (s7) at (0,-4.8) {\bfseries Done in $<$2 min --- no context switch};

  \foreach \a/\b in {s1/s2,s2/s3,s3/s4,s4/s5,s5/s6,s6/s7}
    \draw[arr] (\a) -- (\b);

  \node[note, right=0.5cm of s3, text width=1.6cm, align=center]
    {Gemini 2.5 Pro\\ context-aware};
  \node[note, right=0.5cm of s5, text width=1.8cm, align=center]
    {Semantic Scholar\\ + CrossRef};
\end{tikzpicture}
\caption{\textbf{Smart citation workflow.} From a blank template to a fully
  cited draft paragraph in under two minutes, entirely within Bibby, with no
  external tools or tab switching.}
\label{fig:citeworkflow}
\end{figure}
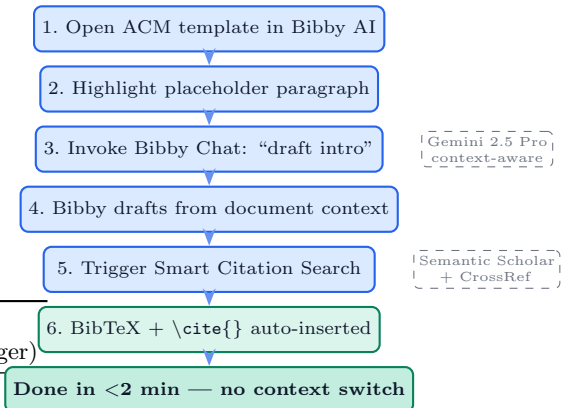

Figure~\ref{fig:citeworkflow} illustrates the citation workflow. A researcher
opens a blank ACM template in Bibby and invokes Bibby Chat: ``draft a
motivation paragraph for a paper on federated learning for clinical NLP.''
Bibby generates a structured paragraph grounded in the full document context.
The researcher triggers Smart Citation Search; Bibby surfaces five
high-relevance references from Semantic Scholar, auto-generates BibTeX, and
inserts \texttt{\textbackslash cite\{\}} commands. Total elapsed time:
under two minutes, with no context switching.

\subsection{Pre-Submission Review and Error Resolution}

\begin{figure}[h]
\centering
\begin{tikzpicture}[font=\scriptsize]

  \draw[rounded corners=4pt, draw=bibbygray, thick, fill=white]
    (-0.1,-0.1) rectangle (6.5,3.4);
  \node[anchor=north west, font=\tiny\bfseries, text=bibbygray]
    at (0.1,3.4) {\LaTeX{} Editor --- main.tex};

  \node[anchor=north west, font=\ttfamily\scriptsize]
    at (0.15,3.0) {\textbackslash usepackage\{booktabs\}};
  \node[anchor=north west, font=\ttfamily\scriptsize]
    at (0.15,2.55) {\textbackslash usepackage\{graphicx\}};

  \fill[warnlight] (0,1.85) rectangle (6.4,2.22);
  \draw[warnred, very thick] (0,1.85) -- (0,2.22);
  \node[anchor=north west, font=\ttfamily\scriptsize, text=warnred]
    at (0.15,2.22)
    {\textbackslash begin\{algorithm\} \% $\leftarrow$ \textcolor{warnred}{error}};

  \node[anchor=north west, font=\ttfamily\scriptsize]
    at (0.15,1.8) {\quad \textbackslash KwIn\{data $x$\}};
  \node[anchor=north west, font=\ttfamily\scriptsize]
    at (0.15,1.35) {\quad \textbackslash Return $f(x)$};
  \node[anchor=north west, font=\ttfamily\scriptsize]
    at (0.15,0.9) {\textbackslash end\{algorithm\}};

  \draw[rounded corners=4pt, draw=bibbyblue, thick, fill=bibbylightblue]
    (-0.1,-2.8) rectangle (6.5,-0.25);

  \node[anchor=north west, font=\scriptsize\bfseries, text=warnred]
    at (0.1,-0.35) {$\blacktriangle$ Compilation Error -- Line 9};
  \node[anchor=north west, text width=5.8cm]
    at (0.1,-0.75)
    {\texttt{\textbackslash algorithm} env.\ requires the \texttt{algorithm2e}
     package, which is not loaded.};
  \node[anchor=north west, text width=5.8cm, font=\itshape]
    at (0.1,-1.35) {Suggested fix:};
  \node[anchor=north west, font=\ttfamily\scriptsize]
    at (0.5,-1.65)
    {\textbackslash usepackage[ruled,vlined]\{algorithm2e\}};

  \draw[rounded corners=3pt, draw=bibbygreen!70!black, thick,
        fill=bibbygreen!25]
    (0.1,-2.65) rectangle (2.8,-2.15);
  \node[font=\scriptsize\bfseries, text=bibbygreen!40!black]
    at (1.45,-2.40) {\checkmark\ Apply Fix (1 click)};

  \draw[rounded corners=3pt, draw=bibbygray, thick, fill=white]
    (3.0,-2.65) rectangle (5.2,-2.15);
  \node[font=\scriptsize, text=bibbygray] at (4.1,-2.40) {Dismiss};

  \draw[-Stealth, thick, draw=warnred]
    (3.2,1.85) -- (3.2,-0.25);

  \node[rectangle, rounded corners=3pt, fill=bibbygreen!20,
        draw=bibbygreen!60, font=\tiny\bfseries, inner sep=3pt]
    at (5.3,-2.90) {Fixed in $<$10 s};

\end{tikzpicture}
\caption{\textbf{Bibby error detection and one-click fix.} Bibby identifies
  the missing \texttt{algorithm2e} package, highlights the offending line,
  explains the error in plain English, and offers a one-click fix that inserts
  the correct \texttt{\textbackslash usepackage} directive.}
\label{fig:errordemo}
\end{figure}
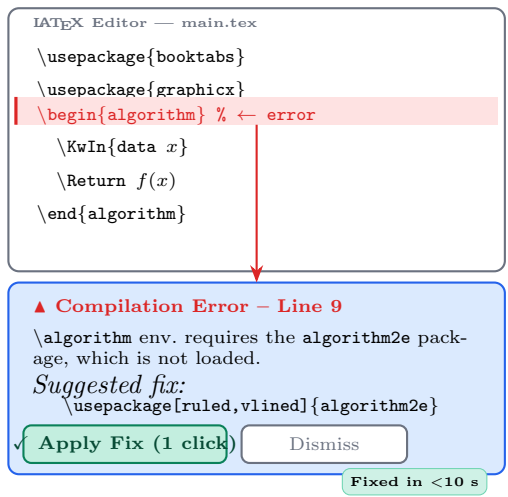

Before submitting to ICML, a researcher uploads their PDF to Bibby's AI Paper
Reviewer, selecting the ICML venue profile. Within 30 seconds, Bibby returns
structured, venue-calibrated feedback. The researcher returns to the editor
and encounters a compilation error. Figure~\ref{fig:errordemo} shows Bibby
highlighting the offending line, explaining in plain English that
\texttt{\textbackslash begin\{algorithm\}} requires the \texttt{algorithm2e}
package, and offering a one-click preamble fix. The entire cycle
completes in under ten seconds.

\section{Privacy and Responsible AI}

A core commitment of Bibby is that user research data is never used to train
AI models. Documents are processed ephemerally; no content is retained beyond
the active inference session. This is especially important for researchers
working under embargo, handling clinical data, or subject to institutional IP
policies. All AI features can be individually enabled or disabled. Bibby
supports ORCID and university SSO for institutional authentication, and GitHub
synchronisation for version-controlled manuscript management.

\section{Related Work}

\mypara{AI-assisted academic writing}
Lee et al.~\cite{lee2024design} identify context-awareness and in-editor
integration as critical unsolved challenges for writing assistants.
Liebling et al.~\cite{liebling2025ai} demonstrate that structured AI feedback
improves submission quality. PEARL~\cite{mysore2024pearl} explores LLM writing
assistant personalisation through generation-calibrated retrieval.

\mypara{In-editor \LaTeX{} assistance}
OverleafCopilot~\cite{wen2024overleafcopilot} augments Overleaf via a browser
extension providing GPT-4-based suggestions. PaperDebugger~\cite{hou2025paperdebugger}
introduces a Kubernetes-native multi-agent architecture for in-editor review
and editing, also implemented as a Chrome extension. Both systems demonstrate
the value of editor-integrated AI but rely on DOM injection, creating fragility
and limiting document-structure access. Bibby eliminates this by building AI
natively into the editor data model.

\mypara{\LaTeX{} error repair}
Prior work on LLM-based program repair~\cite{tian2023repair} is applicable to
markup languages, but existing benchmarks focus on general programming
languages. LaTeXBench-500 is the first benchmark specifically designed for
\LaTeX{} error detection and repair.

\section{Conclusion}

Bibby AI presents a new model for AI-augmented academic writing: rather than
wrapping an existing editor with an external AI layer, Bibby integrates AI
natively into every facet of the research writing workflow. Our evaluation on
LaTeXBench-500 establishes that Bibby's \LaTeX{} error detection (91.4\% DA)
and one-click fix accuracy (83.7\% FA) substantially exceed both Overleaf's
native diagnostics and OpenAI Prism across all six categories. The
eight-feature suite---AI writing assistant, smart citation search,
table/equation generation, paper review, abstract generation, literature
review, deep research, and error auto-fix---makes Bibby the first editor to
address the complete research writing lifecycle within a single,
privacy-preserving interface. Bibby is freely available at
\href{https://trybibby.com}{trybibby.com} (free plan, no credit card required;
Pro from \$15/month).

\end{document}